\documentclass[12pt,preprint]{emulateapj}

\def\etal{\emph{et al. }}
\def\ie{\emph{i.e. }}
\def\cf{\emph{c.f. }}

\def\aposteriori{\emph{a posteriori }}

\shorttitle{GJ 623 precision masses}
\shortauthors{F. Martinache \etal}

\begin{document}

\bibliographystyle{apj}

\title{Precision Masses of the low-mass binary system GJ 623}

\author{Frantz Martinache\altaffilmark{1},
  James P. Lloyd\altaffilmark{1},
  Michael J. Ireland\altaffilmark{2},
  Ryan S. Yamada\altaffilmark{1},
  Peter G. Tuthill\altaffilmark{3}}

\altaffiltext{1}{Department of Astronomy, Cornell University, Ithaca NY}
\altaffiltext{2}{Division of Geological and Planetary Sciences,
  California Institute of Technology, Pasadena CA}
\altaffiltext{3}{School of Physics, University of Sydney, Sydney NSW
  Australia}

\begin{abstract}
We have used Aperture Masking Interferometry and Adaptive Optics (AO)
at the Palomar 200'' to obtain precise mass measurements of the binary
M dwarf GJ 623.
AO observations spread over 3 years combined with a decade of radial
velocity measurements constrain all orbital parameters of the GJ 623
binary system accurately enough to critically challenge the models.
The dynamical masses measured are $m_{1}=0.371\pm0.015\,M_{\sun}$
(4\%) and $m_{2}=0.115\pm0.0023\,M_{\sun}$ (2\%) for the primary and
the secondary respectively.
Models are not consistent with color and mass, requiring very low
metallicities.
\end{abstract}


\section{Introduction}

The mass of a star, along with its metallicity and age, is the
fundamental parameter that determines its position along an
evolutionary track.
Even if binarity, rotation, magnetic fields and other parameters
also affect stellar interiors, what is known as the Vogt-Russel
theorem remains an important rule, and the mass an essential
parameter of stellar evolution.
Binary systems offer the ideal test to infer dynamical masses
independent of the use of a stellar model such as those of
\citet{1998A&A...337..403B}, or an empirical mass-luminosity
(hereafter M/L) relation.
Such relations, available for both visible \citep{1999ApJ...512..864H}
and near infrared \citep{2000A&A...364..217D} are important
astrophysical tools, fairly well constrained for intermediate mass
stars.
However, the Solar neighborhood is dominated by low mass stars, in
both number and total mass \citep{1998ASPC..134...28H},
and as dust condenses in the atmospheres of these cool stars, the
models meet new unknowns.
There also are puzzles in the stellar structure of M dwarfs and the
extrapolation of these models to substellar objects remains untested.
Below $0.6\,M_{\Sun}$, both M/L relations and models will benefit from
model-independent determinations of high precision stellar masses,
which can be achieved by combining radial velocimetry measurements and
high angular resolution imaging \citep{2000A&A...364..665S,
2004ASPC..318..166D}. The complementarity of the two techniques yields
substantial benefits, even in the regime where the time baseline of
the observations is shorter than the orbital period
\citep{2002ApJ...574..426E}.

We have observed known M dwarfs binaries with precise radial
velocity measurements published by \citet{2002ApJS..141..503N}.
These late type M dwarfs are ideal observing targets for imaging with
the PALAO Adaptive Optics (AO) system and the PHARO infrared camera
\citep{2001PASP..113..105H}, optimized for the near infrared bands.
At small angular separation (\ie  $<2\lambda/d$), the sensitivity of
the detection of faint companions can be improved by combining AO with
aperture masking interferometry \citep{2006ApJ...649..389P,
2006astro.ph..7516L}.
This paper provides the astrometry of the binary system GJ 623,
successfully observed with both AO and AO + aperture masking.
Combined with the radial velocity measurement, the astrometry provides
precise dynamical masses (better than 2\%) of the GJ 623 binary
system.
Combined with the J,H,K photometry, this measurement adds new
constraints to the models and M/L relations.

\section{The observations}

\subsection{The primary}

GJ 623 (aka LHS 417, HIP 80346) is a high proper motion M2.5 dwarf 8
pc from the Sun. It is a long known astrometric binary, first
characterized by \citet{1978PASP...90..226L}.
GJ 623 has proven to be a ideal test for different observational
techniques, such as radial velocimetry \citep{1989ApJ...341..961M,
2002ApJS..141..503N} and speckle interferometry
\citep{1987ApJ...319L..93M}.
This binary system has also been directly imaged in the visible with
the COSTAR corrected HST/FOC \citep{1996A&A...315..418B}.

Our observations of GJ 623 were performed with the PHARO instrument on
the Palomar 200'' telescope and with the NIRC2 instrument on the Keck
II telescope.
The companion of GJ 623 was detected in six observing runs: September
2003, June 2004, January 2005 and February 2006, at
Palomar, using J, H and K$_s$ filters, and June and August 2006 at
Keck, with the $1.58\,\mu m$ H$_{cont}$ filter.
Tables \ref{tbl:astro} and \ref{tbl:photo} respectively gather the
astrometric and photometric measurements made at Palomar and Keck.

Four of our observations consist of conventional direct imaging with
AO. These images were dark substracted, flat fielded and analysed with
a custom IDL program, using the latest PHARO and NIRC2 plate scale and
orientation characterization by \citet{2004ApJ...617.1330M}.
The location of the companion (angular separation and position angle)
and the constrast ratio are precisely determined with a cross
correlation of the images.
Not surprisingly, doubling the diameter of the telescope (\cf the Keck
data points) significantly improves the precision of the measurements
and roughly reduces the error bars by a factor 3.

The data also include observations in aperture masking interferometry
with AO. This technique \citep{2000PASP..112..555T}, recently
described by \citet{2006astro.ph..7516L}, is appropriate to 
the detection of faint companion at small angular separation
(typically less than 2 $\lambda/D$), where direct AO imaging has so
far proven difficult.

Indeed, present AO systems focus on achieving high contrast at
moderate angular separation $>4\lambda/D$.
Below this limit, the variance of the speckle background dominates the
photon noise by several orders of magnitude \citep{1999PASP..111..587R}.
This prevents us from calibrating the Point Spread Function (PSF),
with a precision sufficient to discriminate faint companions from the
bright star's residual speckes.

This issue can be evaded by using interferometric techniques. By
sampling a few spatial frequencies only, a mask located in the pupil
plane permits us to completely decompose the PSF into a finite set of
Fourier components. Non-redundancy of the baselines passed by the mask
ensures that each frequency is sampled only once and the visibilities
can be used to form closure phases
\citep{1986Natur.320..595B,1988AJ.....95.1278R,1989AJ.....97.1510N}.
This observable rejects both atmospheric noise and calibration errors
of the wavefront sensor.
The only drawback is the tranmission of such a mask: between
5 and 15 \% for the ones used at Palomar. However, in the speckle
noise limited regime, light loss does not result in a loss of
sensitivity.

Our analysis also uses a HST observation dating to 1994, and
originally published by \cite{1996A&A...315..418B}: an angular
distance of $330 \pm 20$ mas and a position angle of $7.0 \pm
2.6^{\circ}$ on June 11, 1994.
Even though this is inferior to the precision achieved with Palomar
and Keck, which are much larger telescopes, the fact that this
observation was performed more than 10 years ago adds an important
constraint on the period of the binary.

\begin{table}[htbp]
\hfill 
\begin{tabular}{| c | c | c | c @{ $\pm$ } c | c @{ $\pm$ } c |}
   \hline
   \textbf{Julian Date} & \textbf{Band} & \textbf{Pupil} &
   \multicolumn{2}{|c|}{\textbf{Sep.}} & \multicolumn{2}{|c|}{\textbf{PA}} \\
   \textbf{(-2450000)} & & (see text) &
   \multicolumn{2}{|c|}{(mas)} & \multicolumn{2}{|c|}{(deg)} \\ \hline
   2896.6 & Ks        & 9 hole   & 240.4 & 9.7  &  79.31 & 2.0  \\ \hline
   3163.8 & [Fe II]   & 18 hole  & 340.5 & 20.4 &  49.16 & 4.3  \\ \hline
   3402.1 & $K_s$     & full ap. & 350.7 & 2.2  &  28.66 & 0.5  \\ \hline
   3780.0 & J,H,K$_s$ & full ap. & 257.1 & 3.1  & -12.76 & 0.6  \\ \hline
   3909.5 & H$_{cont}$ & Nirc2    & 176.7 & 1.1  & 318.2  & 0.2  \\ \hline
   3962.3 & H$_{cont}$ & Nirc2    & 138.7 & 0.9  & 295.4  & 0.2  \\ \hline
\end{tabular}
\hspace*{\fill}
\caption{
  Astrometric measurements at Palomar and Keck: angular separation
  and position angle of GJ 623 B.}
\label{tbl:astro}
\end{table}

\begin{table}[htbp]
\hfill
\begin{tabular}{| c | c | r @{ $\pm$ } l|}
  \hline
  \textbf{Julian Date} & \textbf{Filter} &
  \multicolumn{2}{|c|}{\textbf{$\Delta mag$}}    \\
  (-2450000)  &       & \multicolumn{2}{|c|}{ } \\ \hline
  2896.6 & K$_s$      & 3.051 & 0.826 \\ \hline
  3163.8 & Fe II      & 2.903 & 0.476 \\ \hline
  3402.1 & K$_s$      & 2.604 & 0.047 \\ \hline
  3780.0 & J          & 2.691 & 0.038 \\
         & H          & 2.860 & 0.039 \\
         & K$_s$      & 2.789 & 0.014 \\ \hline
  3909.5 & H$_{cont}$  & 2.794 & 0.033 \\ \hline
  3962.3 & H$_{cont}$  & 2.781 & 0.016 \\ \hline
\end{tabular}
\hspace*{\fill}
\caption{Photometric measurements of GJ 623 B.}
\label{tbl:photo}
\end{table}

From the 2MASS catalog, the apparent magnitudes of GJ 623 AB are
$J = 6.638 \pm 0.024$, $H = 6.141 \pm 0.021$ and $K = 5.915 \pm
0.023$.

The Hipparcos parallax $\pi = 124.34\pm1.16\,mas$ originally published
by \citet{1997A&A...323L..49P} was based on the radial velocimetry
measurement of \citet{1989ApJ...341..961M}.
The parallax was recalculated by \citet{2005A&A...442..365J} after the
publication of an improved radial velocity curve by
\citet{2002ApJS..141..503N}.
It is this revised parallax $\pi = 125.81\pm1.19\,mas$ (Pourbaix,
D., {\it priv. comm.}) that we adopt here.
The following absolute magnitudes can be deduced for GJ 623 AB:
$M_J = 7.137 \pm 0.052$,
$M_H = 6.640 \pm 0.051$ and
$M_K = 6.414 \pm 0.052$.

\section{Method: extracting orbital parameters}

The starting point in the determination of the orbital parameters of
the GJ 623 system, is the set of radial velocity measurements published
by \cite{2002ApJS..141..503N} and available in the online version of the
{\it Ninth Catalogue of Spectroscopic Binary Orbits}
\citep{2004A&A...424..727P} at \url{http://sb9.astro.ulb.ac.be}.
The parameters deduced from the Keplerian fit to the data are summarized
in table \ref{tbl:data}. They have been directly used by
\cite{2005A&A...442..365J} to  reprocess the {\it Hipparcos Intermediate
Astrometric Data} of \cite{1998A&AS..130..157V} for a complete
characterization of the dynamical elements of the system.

The approach detailed in this paper consists of combining together,
with no a priori assumptions, the orginal radial velocity data and our
astrometric observations.

\subsection{Keplerian Orbits}

One uses the standard 2-body solution to parametrize the location of the
companion orbiting GJ 623 A. To an observing date $t$, one associates an
angle $M$ called the mean anomaly:

\begin{equation}
M(t) = \frac{2\pi}{P}(t-T_P),
\end{equation}

\noindent
where $P$ and $T_P$ respectively represent the orbital period and the
epoch at the periastron passage. In the orbital plane, the $(x,y)$
coordinates can simply be expressed as a function of another angle,
the eccentric anomaly $E$. It is the angle between the direction of the
periastron and the current position of the companion, projected onto the
ellipse's circumscribing circle perpendicularly to the major-axis,
measured at the center of the ellipse (\cf fig. \ref{fig:elli}):

\begin{figure}
\plotone{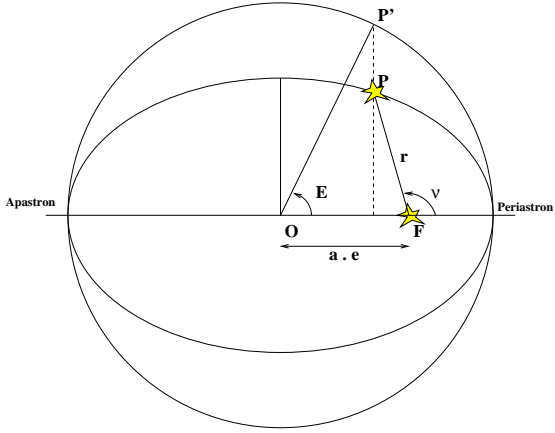}
\caption{Parametrization of an elliptic orbit. Geometric relation
  between the excentric anomaly $E$ and the true anomaly $\nu$.}
\label{fig:elli}
\end{figure}

\begin{eqnarray}
M(E) &=& E-e\sin{E}           \label{eq:kepler} \\
x(E) &=& a (\cos{E}-e)        \label{eq:x}\\
y(E) &=& a\sqrt{1-e^2}\sin{E},\label{eq:y}
\end{eqnarray}

\noindent
where $a$ and $e$ respectively represent the semi-major axis and
the eccentricity of the orbit.
Eq. \ref{eq:kepler}, is called Kepler's equation. One solves this equation
\ie finds the value of $E$ associated with a given $M$ by using the
following classical iterative algorithm:

\begin{eqnarray}
E_0&=&M \nonumber \\
E_{n+1}&=&M+e\sin{E_n}.
\end{eqnarray}

\subsubsection{Radial Velocity Orbital Models}

One can use the standard 2-bodies solution (eq. \ref{eq:x} and \ref{eq:y}) to
compute the coordinates of the primary component in the orbital plane, relative
to the center of mass of the system:

\begin{equation}
\left[
\begin{array}{c}
  x_P(E) \\
  y_p(E)
\end{array}
\right]
 = \frac{m_2}{m_T} \times
\left[
\begin{array}{c}
         x(E) \\
         y(E)
\end{array}
\right],
\end{equation}

\noindent
where $m_T$ and $m_2$ respectively stand for the total mass (primary +
secondary) and the mass of the secondary.
In the orbital plane, the velocity vector can be calculated with the
following partial derivative:

\begin{equation}
V = \frac{\partial}{\partial t}
\left[
\begin{array}{c}
         x\\
         y
\end{array}
\right]
= \frac{\partial}{\partial E}
\left[
\begin{array}{c}
         x\\
         y
\end{array}
\right]
\times \frac{\partial E}{\partial M}
\times \frac{\partial M}{\partial t}.
\end{equation}

This derivation leads to the following $x$ and $y$ components of the
velocity:

\begin{equation}
\left[
\begin{array}{c}
  v_x(E) \\
  v_y(E)
\end{array}
\right]
 = \frac{m_2}{m_T} \times \frac{2\pi a}{P(1-e\cos{E})}
\left[
\begin{array}{c}
         -\sin{E} \\
         \sqrt{1-e^2}\cos{E}
\end{array}
\right].
\end{equation}

The Radial Velocity (hereafter RV) is the component of this velocity
projected on the line of sight. Its expression therefore involves
both the argument of the periastron $\omega_0$ and the inclination of
the system $i$. The RV is not sensitive to the orientation of the system
on the sky, \ie to the value of the argument of the ascending node
$\Omega_1$:

\begin{eqnarray}
RV(E,\omega_0, i) = [v_x(E)\sin{\omega_0} + v_y(E)\cos{\omega_0}]\sin{i} + V_0,
\end{eqnarray}

\noindent
where $V_0$ is a constant term, the mean RV.

\subsubsection{Astrometric Orbital Models}

Unlike the case of RV data for which equations are expressed from the
center of mass of the system, the astrometric measurements gathered in
Table \ref{tbl:astro} provide the instant position of the companion
relative to the primary.
In this frame, the trajectory of the companion is an ellipse, whose
focus is the primary component. The parametric equation of this ellipse
is:

\begin{equation}
  r(\nu) = \frac{a (1-\epsilon^2)}{1+\epsilon\cos\nu},
\label{eq:elli}
\end{equation}

\noindent
where $a$ is the semi-major axis, $\epsilon$, the eccentricity
($0<\epsilon<1$) and $\nu$, an angle called the true anomaly, which
is the angle between the direction of the periastron and the current
position of an object on its orbit ($P$ on fig. \ref{fig:elli}),
measured at the focus of the ellipse.

Fig. \ref{fig:elli} illustrates the one-to-one correspondance between
eccentric and true anomaly. The projection of the point P on the major
axis provides the following relation:

\begin{equation}
  r\cos{\nu} = a(\cos{E} - \epsilon).
  \label{eq:truecc}
\end{equation}

Together, eq. \ref{eq:elli} and \ref{eq:truecc} lead to the primary to
secondary distance $r$ and the true anomaly $\nu$, as functions of the
eccentric anomaly $E$:

\begin{eqnarray}
  r &=& a (1-e\cos{E}), \label{eq:r(E)}\\
  \cos{\nu}&=&\frac{\cos{E}-\epsilon}{1-\epsilon\cos{E}}. \label{eq:nu(E)}
\end{eqnarray}

At a given observing date $t$, one needs once more to solve Kepler's
equation (\cf eq. \ref{eq:kepler}) to determine the corresponding
eccentric anomaly $E$. The location of the companion along the orbit is
provided by eq. \ref{eq:r(E)} and \ref{eq:nu(E)}.

Contrary to the radial velocity, which is the component of the velocity
projected on the line of sight, one measures here the position of the
companion projected on the celestial sphere.
Right ascension $\alpha$ and declination $\delta$ of the secondary
(relative to the primary) are given by the following relations:

\begin{eqnarray}
  \alpha &=& r \times [\cos(\nu+\omega_0) \sin{\Omega_1} + \nonumber \\
  & &\sin(\nu+\omega_0) \cos{i} \cos{\Omega_1}]\label{eq:alpha} \\
  \delta &=& r \times [\cos(\nu+\omega_0) \cos{\Omega_1} - \nonumber \\
  & &\sin(\nu+\omega_0) \cos{i} \sin{\Omega_1}].\label{eq:delta}
\end{eqnarray}

\subsection{$\chi^2$ fitting}

We use the model presented in the previous section, to fit a 9
parameter model to 26 observables. The observables are 2 coordinates
for 7 astrometric data points and 12 radial velocities.
The 9 parameters are the 6 orbital elements, \ie the 
semi-major axis $a$, the eccentricity $e$, the longitude of the
ascending node $\omega_0$, the inclination $i$, the argument of
periastron $\Omega_1$ and the orbital period $P$, plus the RV offset
$V_0$ and the semi-amplitude of the RV curve $K_1$. 

In the case of a conventional analysis of RV data, the main observable
is the semi-amplitude $K_1$, which once combined with the period and
the eccentricity of the orbit, provides the mass function:

\begin{equation}
  f(M_1,M_2,i) = \frac{M_2^3\sin{i}^3}{(M_T)^2}.
\end{equation}

This combined analysis RV+astrometry allows us to separate the
geometrical effects from the semi-amplitude. This produces another
composite observable, somewhat simpler than $K_1$, a
``pseudo-amplitude", whose formal expression is:

\begin{equation}
A_1 = \frac{M_2}{M_T}\times\frac{2\pi a}{P}.
\label{eq:pseudo}
\end{equation}

\begin{figure}
\center{\resizebox*{0.5\textwidth}{!}{\includegraphics{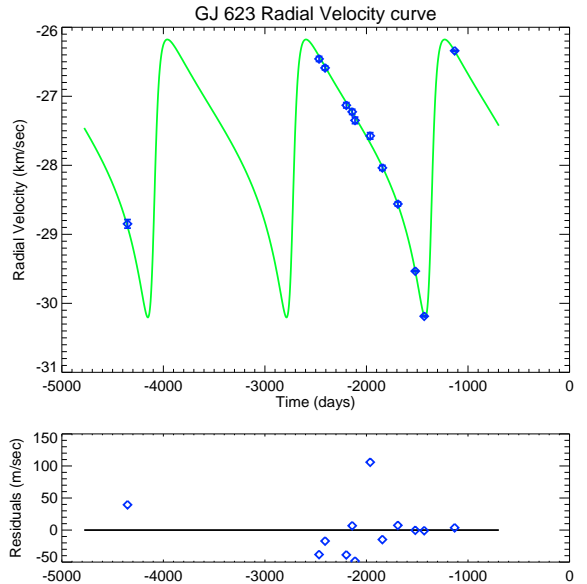}}}
\caption{
Radial Velocity curve of GJ 623A. The measurements and associted uncertainties
reported by \citep{2002ApJS..141..503N} are represented in blue. The green
solid line represents the Keplerian orbit with the parameters derived by
$\chi^2$ fitting.
}
\label{fig:rv}
\end{figure}

\noindent
This is the parameter that will be used to determine the dynamical
masses of the two components of the binary.
\begin{figure}
\center{\resizebox*{0.5\textwidth}{!}{\includegraphics{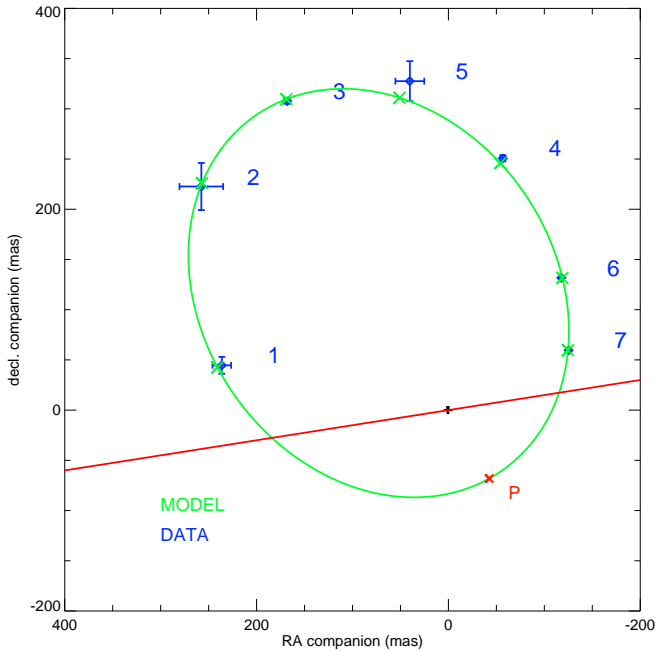}}}
\caption{Orbit of GJ 623B. Measurements and associated uncertainties 
are represented in blue. Points 1-4: PHARO observations, point 5: Hubble
observations reported in \citep{1996A&A...315..418B} and points 6-7: Nirc2
observations. In green, the Keplerian fit to the orbit. The red line is the
line of nodes, the point P marks the position of the Periastron.
}
\label{fig:orbit}
\end{figure}
The final result of this 9 parameter fit, with 17 degrees of freedom is
represented on Figure \ref{fig:orbit} for the astrometry and \ref{fig:rv}
for the radial velocimetry.
Our solution exhibits a final reduced $\chi_{\nu}^2 = 1.03$, which is,
despite the heterogeneity of the data, close enough to unity to ensure
confidence in our estimation of the error bars.
It is dominated by the velocimetry, despite having fewer measurements
than the astrometry.

The confidence interval of each of the parameters is determined by
analysis of the likelihood function. If we assume that the the noise
associated to our measurements is Gaussian, and that our parameters
are independent, the likelihood can be approximated by:

\begin{equation}
  L(\textrm{parameters}) \propto exp\bigg(-\frac{\chi^2}{2}\bigg).
\end{equation}
The computation of a 9-dimensional likelihood function requires a lot
of CPU time. This difficulty may be circumvented by confining the
search to a subset of the total space. This is achieved by fixing the
values of certain parameters and calculating the joined likelihood for
the remaining parameters.
However, one needs to check \aposteriori that the validity of the
hypothesis of independent parameters.

\begin{figure*}
\center{\resizebox*{0.95\textwidth}{!}{
\includegraphics{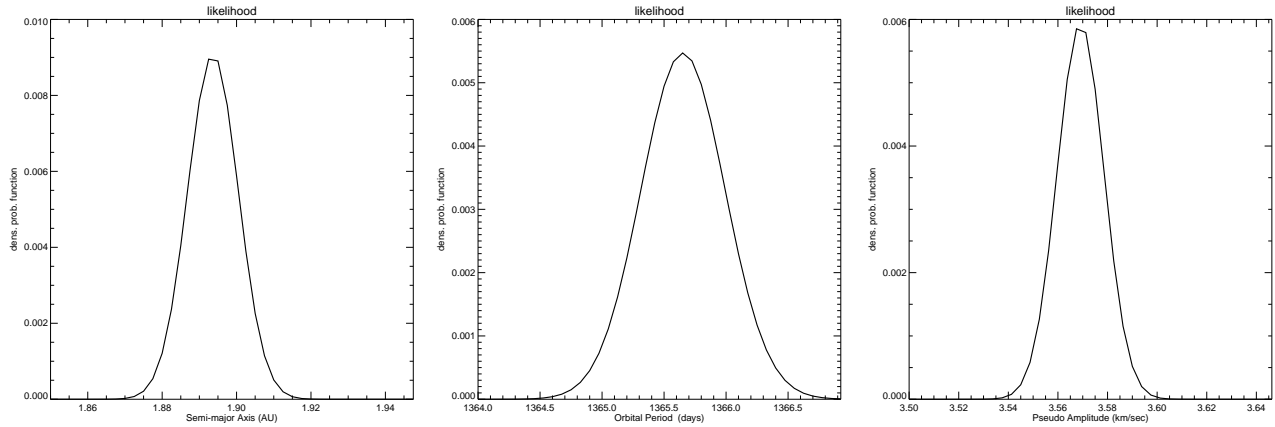}}}
\caption{
  Likelihood functions of semi-major axis $a$, period $P$ and radial
  velocity pseudo-amplitude $A_1$, used to determine the dynamical
  masses of the GJ 623 system.
}
\label{fig:likely}
\end{figure*}

This analysis reveals that among the 9 parameters, only the argument
of the periastron $\omega_0$ and the longitude of the ascending node
$\Omega_1$ exhibit significant correlation.
These two parameters are constrained within $0.5^{\circ}$
(\cf Table \ref{tbl:data}). The hypothesis of independent parameters
is valid, as expected from our good coverage in both radial
velocimetry and astrometry. The uncertainty associated to each
parameter is taken equal to the standard deviation of its associated
likelihood function.
The Gaussian-like likelihoods of the parameters used in the
determination of the dynamical masses are shown on
Fig. \ref{fig:likely}.

\section{Characteristics of GJ 623 AB}

The orbital parameters and the confidence intervals derived from the
likelihood analysis are summarized in Table \ref{tbl:data}. This
combined analysis significantly improved the constraints on the epoch
at periastron passage, from $T_P=2451298\pm10$
\citep{2002ApJS..141..503N} to the new value $T_P=2451313.3\pm0.6$.
The RMS of the Radial Velocity fit is $56$ m/s for a semi-amplitude
of $K_1 = 2.01\pm0.01$ km/s, which is comparable to the original
fit (RMS=51 m/s for $K_1 = 2.08\pm0.04$ km/s).

\begin{table}[htbp]
\hfill
\begin{tabular}{| l | r @{$\pm$} l | r @{$\pm$} l |}
  \hline
   \multicolumn{1}{|c|}{\textbf{parameter}} &
   \multicolumn{2}{|c|}{\textbf{Nidever}} &
   \multicolumn{2}{|c|}{\textbf{This work}}  \\ \hline
   $\Pi$ (mas)        & \multicolumn{2}{|c|}{ } & 125.81  & 1.19     \\ \hline
   $a $ (AU)          & \multicolumn{2}{|c|}{ } & 1.894   & 0.019    \\
   $\alpha$ (mas)     & \multicolumn{2}{|c|}{ } & 237.28  & 0.88     \\ \hline
   $e $               & 0.67 & 0.01             & 0.631   & 0.002    \\ \hline
   $i$  (deg)         & \multicolumn{2}{|c|}{ } & 154.0   & 0.1$^{\circ}$ \\ \hline
   $\Omega_1$         & \multicolumn{2}{|c|}{ } & 98.5    & 0.47$^{\circ}$\\ \hline
   $\omega_o$         & 251     & $1^{\circ}$   & 248.68  &0.46$^{\circ}$\\ \hline
   $P$ (days)         & 1366.1  & 0.4           & 1365.6  & 0.3      \\ \hline
   $T_P$ (reduced JD) & 1298    & 10            & 1313.3  & 0.6      \\ \hline
   $V_0$ (km/s)       & -27.654 & 0.3           & -27.729 & 0.005    \\ \hline
   $A_1$ (km/s)       & \multicolumn{2}{|c|}{ } & 3.57    & 0.01     \\ \hline
\end{tabular}
\hspace*{\fill}
\caption{Orbital Elements}
\label{tbl:data}
\end{table}

\subsection{Dynamical masses}

From the data summarized in Table \ref{tbl:data}, one can determine
the dynamical masses of both components of the system.
The revised parallax figuring in the table takes the binarity into
account.
The semi-major axis $a$, expressed in AU, once combined with
the orbital period, gives the total mass $M_T$ and the associated
uncertainty $\sigma_{T}$:

\begin{eqnarray}
M_T &=& a^3/P^2 \\
\sigma_T/M_T &=& \sqrt{(3\sigma_a/a)^2+ (2\sigma_P/P)^2}.
\end{eqnarray}

The likelihood analysis also determines the ``pseudo-amplitude"
defined by equation \ref{eq:pseudo} and its associated uncertainty.

This composite parameter, combined with the total mass, allows 
independent determination of the mass of the secondary $M_2$:

\begin{eqnarray}
  M_2 &=& M_T \times \frac{A_1\times P}{2\pi a} =
          \frac{A_1\times a^2}{2\pi P}, \\
  \sigma_2/M_2 &=& \sqrt{
    (\sigma_{A_1}/A_1)^2 + (\sigma_P/P)^2 + (2\sigma_a/a)^2
  },
\end{eqnarray}

\bigskip
\noindent
and the mass ratio $M_2/M_T$:

\begin{eqnarray}
  M_R &=& \frac{m_2}{M_T} = \frac{A_1\times P}{2\pi a}, \\
  \sigma_R/M_R &=& \sqrt{(\sigma_{A_1}/A_1)^2 + (\sigma_P/P)^2 +
    (\sigma_a/a)^2}.
\end{eqnarray}

\noindent
A similar analysis is possible for $M_1$. Errors on the masses are
dominated by the uncertainty on the semi-major axis, itself dominated
by the error on the Hipparcos parallax.
As a consequence, with a fractional error of 1 \%, the mass ratio is
better constrained than the mass of the secondary.
The dynamical masses of the GJ 623 system are summarized in Table
\ref{tbl:dynaMass}.
\begin{table}[htbp]
\hfill
\begin{tabular}{| l | c @{ = } c @{ $\pm$ } c |}
  \hline
  \multicolumn{1}{|c|}{\textbf{Quantity}} & 
  \multicolumn{3}{|c|}{\textbf{Value}}                 \\ \hline
  Total Mass     & $M_{T} $ & 0.486 &  0.015 $M_{\sun}$ \\ \hline
  Primary Mass   & $M_{1} $ & 0.371 &  0.015 $M_{\sun}$ \\ \hline
  Secondary Mass & $M_{2} $ & 0.115 & 0.0023 $M_{\sun}$ \\ \hline
\end{tabular}
\hspace*{\fill}
\caption{Dynamical Masses}
\label{tbl:dynaMass}
\end{table}
\subsection{Color, Metallicity \& Kinematics}

The multi-wavelength contrast ratios found in Table \ref{tbl:photo}
may be used to decomposed the observed combined magnitude of the
binary system into magnitudes for individual components.
One determines the following absolute magnitudes:
$M_J =  7.224 \pm  0.052$, 
$M_H =  6.719 \pm  0.051$ and
$M_K =  6.495 \pm  0.052$ for GJ 623 A and
$M_J =  9.915 \pm  0.065$,
$M_H =  9.512 \pm  0.052$ and 
$M_K =  9.269 \pm  0.053$  for GJ 623 B.

With $J-K$ color indices of $0.729 \pm 0.074$ and $0.646 \pm 0.084$
respectively for the primary and the secondary, the GJ 623 AB system
is bluer than the \citet{2000A&A...364..217D} empirical M/L relation.
Therefore, we suspect GJ 623 is of low metallicity.
From the Hipparcos proper motion $(1145.38, -452.37)$ mas/yr
\citep{1997A&A...323L..49P} and the RV offset $V_0 = -27.4$ km/s
determined from our model fit, one can calculate the Galactic space
velocity $(U,V,W) = (-33, 14, -41)$ km/s after correction for
standard solar motion. \footnote{The sign convention is the one of the
IDL astrolib gal\_uvw procedure. Note that the literature is
confusing on this matter. For instance, \cite{1987ApJ...319L..93M}
provide numbers of comparable magnitude but with opposite sign for
all (U,V,W) components.}
This velocity is consistent with an old disk population, and therefore
of subsolar metallicity \citep{2000AJ....119.2843C}, which is
consistent with both components being bluer than the average field
object.

\begin{figure*}
\center{\resizebox*{0.95\textwidth}{!}{\includegraphics{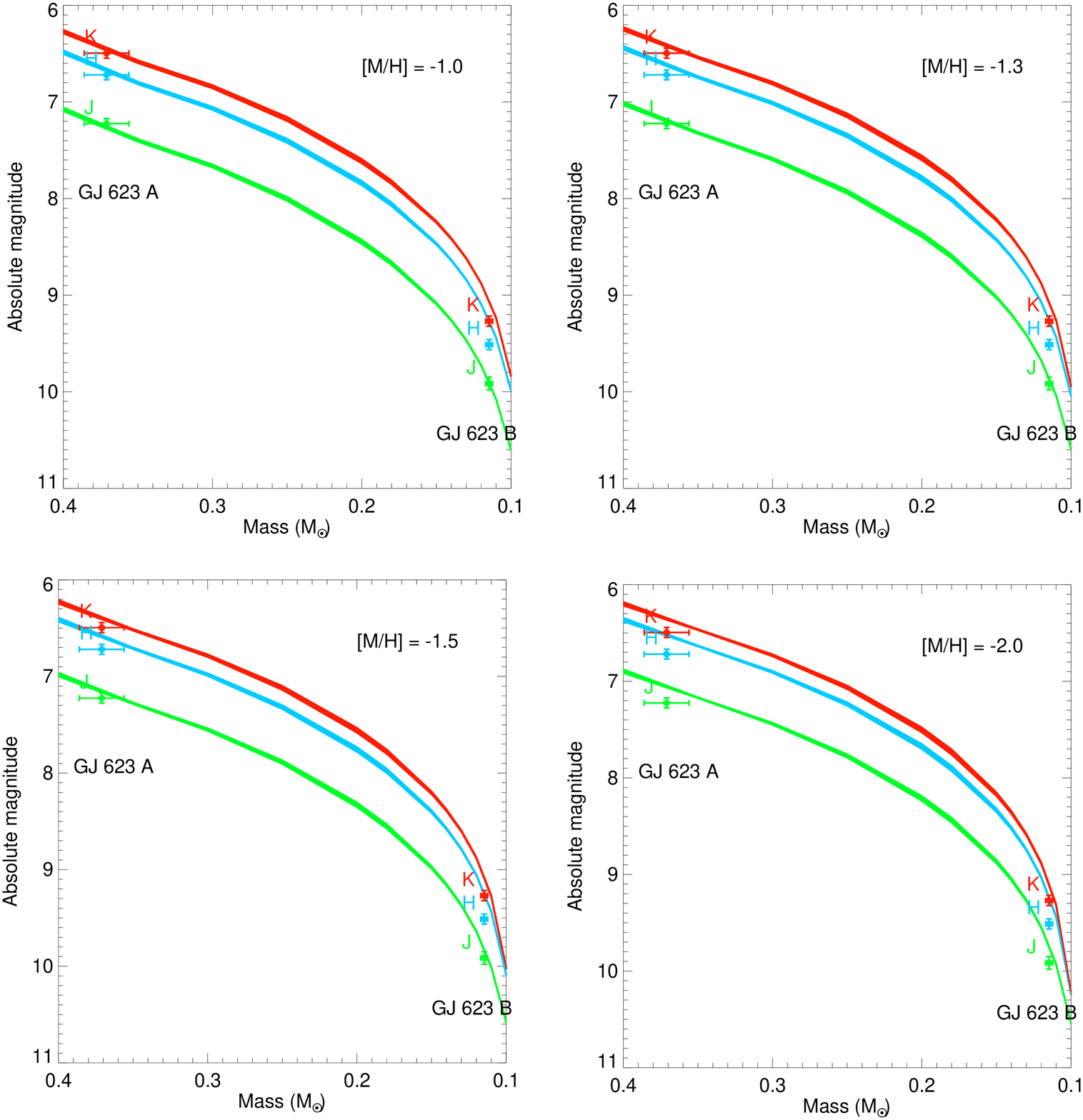}}}
\caption{
  Mass-luminosity diagram for the GJ 623 system in J, H and K bands.
  Comparison with the low mass population II stars models by
  \citet{2000A&A...360..935M} in the 0.1-0.4 solar mass range, for
  different metallicities.
}
\label{fig:MLR}
\end{figure*}

Figure \ref{fig:MLR} compares the location of both components of the
GJ 623 system in a mass-luminosity diagram to the low mass population
II models of \citet{2000A&A...360..935M} for different metallicities.
The trend we observe with these models (\cf Fig. \ref{fig:MLR})
supports GJ 623 AB beeing of subsolar metallicity.
The model that best matches our measurements for the primary is for
$[M/H]=-1.0$. 
For the secondary, the best model predicts a slightly lower mass of
$0.110\pm0.001$ M$_{\sun}$ for $[M/H]=-2.0$.
This large discrepancy in metallicity is inconsistent with the assumption
of a co-eval binary.
The very low metallicity for the secondary 
would be consistent with GJ 623 B belonging to the galactic halo,
which is unlikely according to the kinematics.
Therefore, we conclude that the models do not adequately fit the
data.
\section{Conclusion}

The observation of binary systems is the only way to measure unbiased
masses. As shown in this paper, combined with excellent radial
velocity measurements, a few high angular resolution images provide
sufficient information to constrain the range of possible masses below
the 5\% precision that is required to seriously challenge the models
at the low end of the main sequence.

The application of precision RV methods developed for planet searches,
combined with AO will provide a complete characterization of the
stellar structure of the lower main sequence.

\acknowledgments
We thank the staff and telescope operators of Palomar Observatory and
Keck Observatory for their support. F.M. thanks Terry Herter for his
help.
This work is partially funded by the National Science Foundation under
grants AST-0335695 and AST-0506588.
This publication makes use of the Simbad database, operated at CDS,
Strasbourg, France and the data products from the Two Micron All Sky
Survey, which is a joint project of the University of Massachusetts and
the Infrared Processing and Analysis Center/California Institute of
Technology, funded by the National Aeronautics and Space Administration
and the National Science Foundation.
We wish to extend special thanks to those of Hawaiian ancestry on
whose sacred mountain we are privileged to be guests. Without their
generous hospitality, the observations presented herein would not have
been possible.

\bibliography{biblio}

\begin{thebibliography}{27}
\expandafter\ifx\csname natexlab\endcsname\relax\def\natexlab#1{#1}\fi

\bibitem[{{Baldwin} {et~al.}(1986){Baldwin}, {Haniff}, {Mackay}, \&
  {Warner}}]{1986Natur.320..595B}
{Baldwin}, J.~E., {Haniff}, C.~A., {Mackay}, C.~D., \& {Warner}, P.~J. 1986,
  \nat, 320, 595

\bibitem[{{Baraffe} {et~al.}(1998){Baraffe}, {Chabrier}, {Allard}, \&
  {Hauschildt}}]{1998A&A...337..403B}
{Baraffe}, I., {Chabrier}, G., {Allard}, F., \& {Hauschildt}, P.~H. 1998, \aap,
  337, 403

\bibitem[{{Barbieri} {et~al.}(1996){Barbieri}, {Demarchi}, {Nota}, {Corrain},
  {Hack}, {Ragazzoni}, \& {Macchetto}}]{1996A&A...315..418B}
{Barbieri}, C., {Demarchi}, G., {Nota}, A., {Corrain}, G., {Hack}, W.,
  {Ragazzoni}, R., \& {Macchetto}, D. 1996, \aap, 315, 418

\bibitem[{{Chiba} \& {Beers}(2000)}]{2000AJ....119.2843C}
{Chiba}, M. \& {Beers}, T.~C. 2000, \aj, 119, 2843

\bibitem[{{Delfosse} {et~al.}(2004){Delfosse}, {Beuzit}, {Marchal}, {Bonfils},
  {Perrier}, {S{\'e}gransan}, {Udry}, {Mayor}, \&
  {Forveille}}]{2004ASPC..318..166D}
{Delfosse}, X., {Beuzit}, J.-L., {Marchal}, L., {Bonfils}, X., {Perrier}, C.,
  {S{\'e}gransan}, D., {Udry}, S., {Mayor}, M., \& {Forveille}, T. 2004, in ASP
  Conf. Ser. 318: Spectroscopically and Spatially Resolving the Components of
  the Close Binary Stars, ed. R.~W. {Hilditch}, H.~{Hensberge}, \&
  K.~{Pavlovski}, 166--174

\bibitem[{{Delfosse} {et~al.}(2000){Delfosse}, {Forveille}, {S{\'e}gransan},
  {Beuzit}, {Udry}, {Perrier}, \& {Mayor}}]{2000A&A...364..217D}
{Delfosse}, X., {Forveille}, T., {S{\'e}gransan}, D., {Beuzit}, J.-L., {Udry},
  S., {Perrier}, C., \& {Mayor}, M. 2000, \aap, 364, 217

\bibitem[{{Eisner} \& {Kulkarni}(2002)}]{2002ApJ...574..426E}
{Eisner}, J.~A. \& {Kulkarni}, S.~R. 2002, \apj, 574, 426

\bibitem[{{Hayward} {et~al.}(2001){Hayward}, {Brandl}, {Pirger}, {Blacken},
  {Gull}, {Schoenwald}, \& {Houck}}]{2001PASP..113..105H}
{Hayward}, T.~L., {Brandl}, B., {Pirger}, B., {Blacken}, C., {Gull}, G.~E.,
  {Schoenwald}, J., \& {Houck}, J.~R. 2001, \pasp, 113, 105

\bibitem[{{Henry}(1998)}]{1998ASPC..134...28H}
{Henry}, T.~J. 1998, in ASP Conf. Ser. 134: Brown Dwarfs and Extrasolar
  Planets, ed. R.~{Rebolo}, E.~L. {Martin}, \& M.~R. {Zapatero Osorio}, 28--+

\bibitem[{{Henry} {et~al.}(1999){Henry}, {Franz}, {Wasserman}, {Benedict},
  {Shelus}, {Ianna}, {Kirkpatrick}, \& {McCarthy}}]{1999ApJ...512..864H}
{Henry}, T.~J., {Franz}, O.~G., {Wasserman}, L.~H., {Benedict}, G.~F.,
  {Shelus}, P.~J., {Ianna}, P.~A., {Kirkpatrick}, J.~D., \& {McCarthy}, Jr.,
  D.~W. 1999, \apj, 512, 864

\bibitem[{{Jancart} {et~al.}(2005){Jancart}, {Jorissen}, {Babusiaux}, \&
  {Pourbaix}}]{2005A&A...442..365J}
{Jancart}, S., {Jorissen}, A., {Babusiaux}, C., \& {Pourbaix}, D. 2005, \aap,
  442, 365

\bibitem[{{Lippincott} \& {Borgman}(1978)}]{1978PASP...90..226L}
{Lippincott}, L.~S. \& {Borgman}, E.~R. 1978, \pasp, 90, 226

\bibitem[{{Lloyd} {et~al.}(2006){Lloyd}, {Martinache}, {Ireland}, {Monnier},
  {Pravdo}, {Shaklan}, \& {Tuthill}}]{2006astro.ph..7516L}
{Lloyd}, J.~P., {Martinache}, F., {Ireland}, M.~J., {Monnier}, J.~D., {Pravdo},
  S.~H., {Shaklan}, S.~B., \& {Tuthill}, P.~G. 2006, ArXiv Astrophysics
  e-prints

\bibitem[{{Marcy} \& {Moore}(1989)}]{1989ApJ...341..961M}
{Marcy}, G.~W. \& {Moore}, D. 1989, \apj, 341, 961

\bibitem[{{McCarthy} \& {Henry}(1987)}]{1987ApJ...319L..93M}
{McCarthy}, Jr., D.~W. \& {Henry}, T.~J. 1987, \apjl, 319, L93

\bibitem[{{Metchev} \& {Hillenbrand}(2004)}]{2004ApJ...617.1330M}
{Metchev}, S.~A. \& {Hillenbrand}, L.~A. 2004, \apj, 617, 1330

\bibitem[{{Montalb{\'a}n} {et~al.}(2000){Montalb{\'a}n}, {D'Antona}, \&
  {Mazzitelli}}]{2000A&A...360..935M}
{Montalb{\'a}n}, J., {D'Antona}, F., \& {Mazzitelli}, I. 2000, \aap, 360, 935

\bibitem[{{Nakajima} {et~al.}(1989){Nakajima}, {Kulkarni}, {Gorham}, {Ghez},
  {Neugebauer}, {Oke}, {Prince}, \& {Readhead}}]{1989AJ.....97.1510N}
{Nakajima}, T., {Kulkarni}, S.~R., {Gorham}, P.~W., {Ghez}, A.~M.,
  {Neugebauer}, G., {Oke}, J.~B., {Prince}, T.~A., \& {Readhead}, A.~C.~S.
  1989, \aj, 97, 1510

\bibitem[{{Nidever} {et~al.}(2002){Nidever}, {Marcy}, {Butler}, {Fischer}, \&
  {Vogt}}]{2002ApJS..141..503N}
{Nidever}, D.~L., {Marcy}, G.~W., {Butler}, R.~P., {Fischer}, D.~A., \& {Vogt},
  S.~S. 2002, \apjs, 141, 503

\bibitem[{{Perryman} {et~al.}(1997){Perryman}, {Lindegren}, {Kovalevsky},
  {Hoeg}, {Bastian}, {Bernacca}, {Cr{\'e}z{\'e}}, {Donati}, {Grenon}, {van
  Leeuwen}, {van der Marel}, {Mignard}, {Murray}, {Le Poole}, {Schrijver},
  {Turon}, {Arenou}, {Froeschl{\'e}}, \& {Petersen}}]{1997A&A...323L..49P}
{Perryman}, M.~A.~C., {Lindegren}, L., {Kovalevsky}, J., {Hoeg}, E., {Bastian},
  U., {Bernacca}, P.~L., {Cr{\'e}z{\'e}}, M., {Donati}, F., {Grenon}, M., {van
  Leeuwen}, F., {van der Marel}, H., {Mignard}, F., {Murray}, C.~A., {Le
  Poole}, R.~S., {Schrijver}, H., {Turon}, C., {Arenou}, F., {Froeschl{\'e}},
  M., \& {Petersen}, C.~S. 1997, \aap, 323, L49

\bibitem[{{Pourbaix} {et~al.}(2004){Pourbaix}, {Tokovinin}, {Batten}, {Fekel},
  {Hartkopf}, {Levato}, {Morrell}, {Torres}, \& {Udry}}]{2004A&A...424..727P}
{Pourbaix}, D., {Tokovinin}, A.~A., {Batten}, A.~H., {Fekel}, F.~C.,
  {Hartkopf}, W.~I., {Levato}, H., {Morrell}, N.~I., {Torres}, G., \& {Udry},
  S. 2004, \aap, 424, 727

\bibitem[{{Pravdo} {et~al.}(2006){Pravdo}, {Shaklan}, {Wiktorowicz},
  {Kulkarni}, {Lloyd}, {Martinache}, {Tuthill}, \&
  {Ireland}}]{2006ApJ...649..389P}
{Pravdo}, S.~H., {Shaklan}, S.~B., {Wiktorowicz}, S.~J., {Kulkarni}, S.,
  {Lloyd}, J.~P., {Martinache}, F., {Tuthill}, P.~G., \& {Ireland}, M.~J. 2006,
  \apj, 649, 389

\bibitem[{{Racine} {et~al.}(1999){Racine}, {Walker}, {Nadeau}, {Doyon}, \&
  {Marois}}]{1999PASP..111..587R}
{Racine}, R., {Walker}, G.~A.~H., {Nadeau}, D., {Doyon}, R., \& {Marois}, C.
  1999, \pasp, 111, 587

\bibitem[{{Readhead} {et~al.}(1988){Readhead}, {Nakajima}, {Pearson},
  {Neugebauer}, {Oke}, \& {Sargent}}]{1988AJ.....95.1278R}
{Readhead}, A.~C.~S., {Nakajima}, T.~S., {Pearson}, T.~J., {Neugebauer}, G.,
  {Oke}, J.~B., \& {Sargent}, W.~L.~W. 1988, \aj, 95, 1278

\bibitem[{{S{\'e}gransan} {et~al.}(2000){S{\'e}gransan}, {Delfosse},
  {Forveille}, {Beuzit}, {Udry}, {Perrier}, \& {Mayor}}]{2000A&A...364..665S}
{S{\'e}gransan}, D., {Delfosse}, X., {Forveille}, T., {Beuzit}, J.-L., {Udry},
  S., {Perrier}, C., \& {Mayor}, M. 2000, \aap, 364, 665

\bibitem[{{Tuthill} {et~al.}(2000){Tuthill}, {Monnier}, {Danchi}, {Wishnow}, \&
  {Haniff}}]{2000PASP..112..555T}
{Tuthill}, P.~G., {Monnier}, J.~D., {Danchi}, W.~C., {Wishnow}, E.~H., \&
  {Haniff}, C.~A. 2000, \pasp, 112, 555

\bibitem[{{van Leeuwen} \& {Evans}(1998)}]{1998A&AS..130..157V}
{van Leeuwen}, F. \& {Evans}, D.~W. 1998, \aaps, 130, 157

\end{thebibliography}

\end{document}